\documentstyle[twocolumn,aps,floats]{revtex}

\input psfig.sty
 
\begin{document}
\draft
 
\twocolumn[\hsize\textwidth\columnwidth\hsize\csname %
@twocolumnfalse\endcsname

\bibliographystyle{revtex}


\title{Computation of dendritic microstructures using a level set method} 

\author{Yung-Tae Kim$^{1}$, Nigel Goldenfeld$^{1}$, and Jonathan
Dantzig$^{2}$ }

\address{ $^1$University of Illinois at Urbana-Champaign, Department of
Physics \\1110 West Green Street, Urbana, IL, 61801}

\address{
$^2$University of Illinois at Urbana-Champaign, Department 
of Mechanical and Industrial Engineering 
\\1206 West Green Street, Urbana, IL, 61801
}

\date{\today}
  
\maketitle

\begin{abstract}  

We compute time-dependent solutions of the sharp-interface model of
dendritic solidification in two dimensions by using a level set method. 
The steady-state results are in agreement with solvability theory. 
Solutions obtained from the level set algorithm are compared with
dendritic growth simulations performed using a phase-field model and the
two methods are found to give equivalent results. Furthermore, we perform
simulations with unequal diffusivities in the solid and liquid phases and
find reasonable agreement with the available theory.

\end{abstract}
\vspace{0.2in}

]

Various numerical approaches \cite{Alm93a,Roosen94,Ihle94,Juric96} have
been developed to solve the difficult moving boundary problem that governs
the growth of dendrites \cite{Gli84,Lan87a,Kes88}. Unfortunately, the
direct solution of the time-dependent Stefan problem is troublesome and
usually requires front tracking and lattice deformation in order to
contain the moving solid-liquid interface, which is often very complicated
topologically. In general, the methods developed to tackle the
free-boundary problem have difficulty in handling topology changes, such
as the merging and breaking of surfaces, and are usually not easily
extendible to higher dimensions.

In order to avoid the difficulties associated with tracking a sharp
interface, the phase-field model of solidification has been developed and
is currently the most popular technique for simulating dendritic growth.
The phase-field model avoids the computational difficulties associated
with front tracking by introducing an auxiliary order parameter, or
phase-field, $\psi ({\bf r},t)$ that couples to the evolution of the
thermal field.  The dynamics of $\psi ({\bf r},t)$ are designed to follow
the evolving solidification front \cite{Col85,Cag86a,Kob93,War95},
which is defined by the zero level set $\psi ({\bf r},t)=0$.  Because the
interface is never explicitly tracked, complicated topology changes are
handled easily.  Furthermore, the extension of the phase-field model to
higher dimensions is straightforward.

Although phase-field models have been very useful in studying
solidification patterns, there are still some limitations in this
approach.  The proper use of these models requires that an asymptotic
analysis be performed in order to obtain a mapping between the parameters
of the phase-field equations and the sharp-interface equations
\cite{Lan86a,Cag92,Kar95}.  The asymptotics involve expanding the
phase-field equations in some small parameter proportional to the
interface width, $W$, and as a result, the phase-field model only
reproduces the dynamics of the sharp-interface equations in the limit
where the expansion parameter is sufficiently small.  Computationally, the
grid spacing must be small enough to resolve the interfacial region, which
is on the order of $W$.  This restriction is generally not a problem for
the symmetric model of solidification (where the diffusivities in the
solid and liquid phases are assumed to be the same) because it is possible
to have $W$ on the order of the capillary length \cite{Kar95}.  However,
phase-field asymptotics for unequal diffusivities can be problematic
\cite{Alm98}; correction terms that are inconsistent with the
sharp-interface equations are generated and non-monotonic behavior is
required in the interfacial region, which requires extra grid resolution
and hence slower computational performance.  The generalization of the
phase-field approach to handle discontinuous material properties requires
a better understanding of the mapping between the phase-field model and
the sharp-interface formulation in order to avoid problems with properly
resolving the interface.

The level set method is a computational approach that has the capability
of avoiding the above mentioned limitations of front tracking methods and
phase-field models.  This method, first introduced by Osher and Sethian
\cite{Osher88}, is conceptually similar to a phase-field model in that the
solid-liquid interface is represented as the zero contour of a level set
function, $\phi({\bf r},t)$, which has its own equation of motion.  The
movement of the interface is taken care of implicitly through an advection
equation for $\phi({\bf r},t)$.  Thus, topology changes and the extension
of the method to higher dimensions can be handled in a straightforward
manner.  Unlike the phase-field model, there is no arbitrary interface
width introduced in the level set method; the sharp-interface equations
can be solved directly and, as a result, no asymptotics are required. 
Discontinuous material properties can also be dealt with in a simple
manner.

The level set method has been applied to several problems involving moving
boundaries \cite{Merriman94,Sussman94,Hou97}, including solidification.
Prior work on dendritic growth includes an application of the method to a
boundary integral formulation \cite{Sethian92} as well as the direct
solution of the sharp-interface equations \cite{Chen97}. While these
simulations have reproduced the qualitative features of dendrites, as well
as some quantitatively accurate solutions to exactly soluble problems,
some of the simulations of anisotropic dendritic growth were not
necessarily converged \cite{Chen97}. Furthermore, the results were not
compared with theoretical predictions of dendritic growth.

In this paper, we demonstrate that the level set method can be used to
solve the free-boundary problem for solidification to calculate
quantitatively accurate solutions for dendritic growth. We present results
from simulations in two dimensions and show that the solutions converge to
the steady-state predicted by microscopic solvability theory. 
Time-dependent results are also compared with calculations using a
phase-field model and good agreement is found for all times. Furthermore,
we perform simulations with unequal diffusivities (a case which is not yet
possible with phase-field models) and find that the prediction of Barbieri
and Langer \cite{Bar89} provides a fair quantitative fit to our results.

The solidification of a pure substance is described by a free-boundary
problem for the temperature in the solid and liquid phases, and the
position of the interface between them:  
\begin{eqnarray} 
\partial_t{u} &=& D\nabla^2{u} \\ 
V_n &=& (D\partial_n{u})_{_{Solid}} - (D\partial_n{u})_{_{Liquid}} \\
u_i &=& -d(\theta) \kappa - \beta(\theta) V_n 
\end{eqnarray}
The temperature $T$ has been rescaled as a dimensionless thermal field
$u=(T-T_m)/(L/C_p)$, where $T_m$, $L$, and $C_p$ represent the melting
temperature, the latent heat of fusion, and the specific heat at constant
pressure, respectively.  The thermal diffusivity, $D$, can be different in
the solid and liquid phases.  Eq.~2 describes energy conservation at the
solid-liquid interface, where $V_n$ is the local outward normal interface
velocity and $\partial_n$ refers to the outward normal derivative at the
interface.  Finally, Eq.~3 is known as the Gibbs-Thomson condition and
describes the deviation of the interface temperature, $u_i$, from
equilibrium due to the local curvature, $\kappa$, and interface kinetics. 
$d(\theta)=\gamma(\theta){T_mC_p}/{L^2}$ is the anisotropic capillary
length, proportional to the surface tension $\gamma(\theta)$, and
$\beta(\theta)$ is the anisotropic kinetic coefficient.  Here we assume
that $\beta(\theta)=0$ and that the capillary length has the form
$d(\theta) = d_o(1 - 15 \epsilon \cos4 \theta)$, where $\epsilon$ is the
anisotropy strength and $\theta$ is the angle between the local normal
vector at the interface, $\vec{n}$, and the $x$-axis.

We solve the above free-boundary problem by using a level set algorithm,
which involves the following steps:  i) advancing the interface, ii) 
reinitializing the level set function to be a signed distance function,
and iii) solving for the new thermal field.  The general level set method
is described below.  We wish to note that in our simulations we implement
a {\it localized} level set method, described in detail in
Ref.~\cite{Peng98}, in which calculations of $\phi$ are performed only in
a narrow region around the interface. We have not yet made an attempt to
make our algorithm more computationally efficient by using adaptive mesh
refinement.

i) {\it Advancing the interface}. The level set function is defined as the 
signed normal distance from the solid-liquid interface such that
$\phi$ is positive in the liquid phase, negative in the solid phase, and
zero at the interface. $\phi$ satisfies the pure advection equation 
\begin{equation}
\frac{\partial \phi}{\partial t} + F|\nabla{\phi}| = 0 
\label{levelset}
\end{equation} 
Integrating Eq.~4 for one timestep results in moving the contours of
$\phi$ along the directions normal to the interface according to the
velocity field $F$, which varies in space. $F$ is constructed to be an
extension of the interface velocity, $V_n$, such that $F=V_n$ for points
on the interface and the lines of constant $F$ are normal to the
interface. Thus, advecting $\phi$ according to Eq.~\ref{levelset} moves
the front with the correct velocity.

Rather than using a partial differential equation to generate $F$ (as in
Refs.~\cite{Chen97,Peng98}), we construct $F$ in the following manner: 
$\phi$ represents the normal distance from the solidification front, so
the value of $\phi$ at each gridpoint on the computational lattice can be
used to locate a particular point on the interface. If $\vec{x}_g$ is the
location of the gridpoint, the associated point on the interface is at
$\vec{x}_i=\vec{x}_g - \phi\vec{n}$, where the normal vector
$\vec{n}=\nabla\phi/|\nabla\phi|$. The temperature at $\vec{x}_i$ is then
calculated by using Eq.~3; $\theta$ is easily found from $\vec{n}$, and
the curvature, {$\kappa=\nabla \cdot \vec{n}$}, is interpolated at
$\vec{x}_i$ from values of $\kappa$ at neighboring gridpoints. $\vec{n}$
and $\kappa$ are calculated using standard, centered finite difference
approximations to the partial derivatives of $\phi$. Next, values of $u$
are interpolated in both the liquid and solid phases, a distance $\Delta
x$ (the size of the grid spacing) away from $\vec{x}_i$ along the normal
direction. These two interpolated temperatures are used along with $u_i$
to approximate the difference in the normal derivative of $u$ at
$\vec{x}_i$ and thus find $V_n$ (Eq.~3).  Because $\vec{x}_i$ and
$\vec{x}_g$ lie in the same line normal to the interface, the value of $F$
at $\vec{x}_g$ is simply $V_n$. The field $F$ can be determined at all
gridpoints in this way.

After $F$ is known, the interface can be advanced one timestep. For
stability, we discretize Eq.~\ref{levelset} using a 5th order WENO
(weighted essentially non-oscillatory) scheme in space and a 3rd order
Runge-Kutta scheme in time \cite{Jiang97}. However, the overall accuracy
of our algorithm is second order in space and first order in time.

ii) {\it Reinitialization}. After solving Eq.~\ref{levelset} for one
timestep, the level set function will no longer be equal to the distance
away from the interface. It is necessary to {\it reinitialize} $\phi$ to
be a signed distance function. This step is accomplished by solving
\begin{equation}
\frac{\partial \phi}{\partial t} + S(\phi)[|\nabla{\phi}|-1] = 0
\label{reinitialize}
\end{equation}
to steady state. $S(\phi)$ takes on the value $+1$ in the liquid phase,
$-1$ in the solid phase, and is zero at the interface. We typically
iterate Eq.~\ref{reinitialize} three times in order to obtain an accurate
distance function. Like Eq.~\ref{levelset}, this equation is discretized
using a 5th order WENO scheme in space and a 3rd order Runge-Kutta scheme
in time \cite{Jiang97}.

iii) {\it Solving for the new thermal field}. The thermal field is updated
by solving Eq.~1 using a modified Crank-Nicolson scheme. Different
diffusivities in the two phases can be taken into account by simply noting
the sign of the level set function and using the appropriate diffusion
coefficient in the finite difference stencil. Special care has to be taken
for gridpoints near the interface. If $|\phi| \leq \Delta x$, the level
set function is used to determine whether the front intersects the stencil
and, if so, interpolate where the interface crosses the stencil. The
stencil is then modified to take into account the location of the
interface and the Gibbs-Thomson condition.

We compute four-fold symmetric dendrites in a $L \times L$ square box
using the procedure described above. Solidification is initiated by a
small quarter disk of radius $R_o$ in the lower left-hand corner of the
box. The initial level set function is $\phi(x,y) = \sqrt{x^2 + y^2} -
R_o$, where $x$ and $y$ are the usual Cartesian coordinates. The initial
temperature is $u=0$ in the solid and decays exponentially away from the
interface to $u=-\Delta$ as $\vec{x} \rightarrow \infty$, where the
far-field {\it undercooling} is $\Delta=(T_m-T_\infty)/(L/C_p)$ and
$T_\infty$ is the temperature far ahead of the solidification front in the
liquid.

Eqs.~1-3 have been studied extensively to determine the steady state
features of dendritic growth \cite{Lan87a,Kes88}. According to microscopic
solvability theory, these equations admit a family of discrete solutions. 
Only the fastest growing of this set of solutions is stable.  This
solution is the dynamically selected ``operating state'' for the dendrite
and corresponds to a unique tip shape and velocity.  Recent calculations
of dendritic growth using phase-field models have been found to be in good
agreement with the predictions of microscopic solvability theory
\cite{Kar95,Pro98a}. We observe similar agreement with the use of the
level set algorithm and obtain results that are within a few percent of
theoretical predictions.  Figure 1 shows the tip velocity of the dendrite
versus time for computations at undercoolings of $\Delta=0.65$ and $0.55$.
For all of these simulations $D=1$, $d_o=0.5$, $\beta=0$, $R_o=15$, and
$\epsilon=0.05$.  For the $\Delta=0.65$ simulation, $L = 200$, $\Delta x =
0.2$, and the timestep is chosen to be $\Delta t = 0.002$.  For the
$\Delta=0.55$ simulation, $L = 800$, $\Delta x = 0.4$, and $\Delta t =
0.008$. To ensure grid convergence, $\Delta x$ and $\Delta t$ were refined
until the steady-state tip velocity did not vary by more than $2\%$.

\begin{figure}[tbh]
\leavevmode\centering\psfig{file=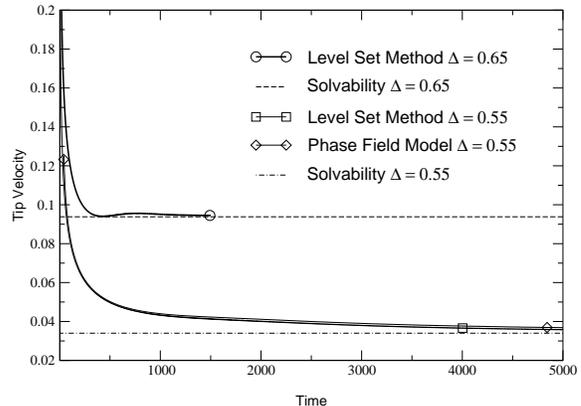,width=\columnwidth,angle=-90}
\caption{Time evolution of the tip velocity for simulations at $\Delta=0.65$ 
and $0.55$.}
\end{figure}

We also compare our level set results with simulations of dendritic growth
performed using a phase-field model.  Although calculations using phase-field
models have been compared with a steady-state theory, there have been no
comparisons made between time-dependent phase-field calculations and
time-dependent solutions of the sharp-interface equations for multi-dimensional
dendritic growth.  The phase-field model calculations presented here were
performed using a special adaptive mesh algorithm, as described in
Ref.~\cite{Pro98a}.  The tip velocity data from the phase-field model and
level set method at $\Delta=0.55$ are in excellent agreement with each other
(within $3\%$), as shown in Figure 1.  Similar agreement is found in the
dendritic shapes for these simulations, presented at time=9400 in Figure 2.
These comparative results, combined with the recent demonstration of the
equivalence of various phase-field models \cite{Kim99}, provide an excellent
foundation for the validity of the phase-field approach in simulating
solidification microstructures.

\begin{figure}[tbh]
\leavevmode\centering\psfig{file=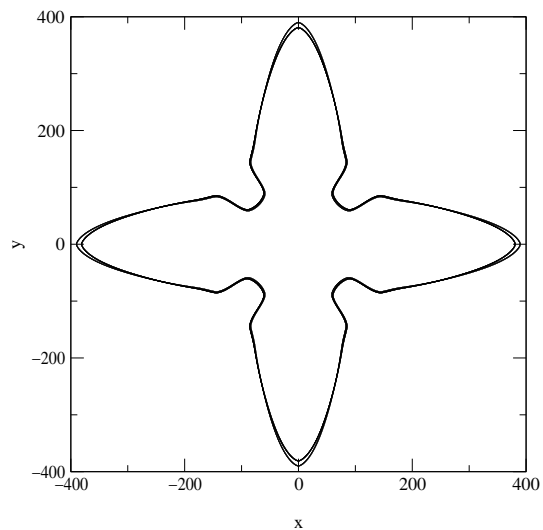,width=\columnwidth}
\caption{Comparison of dendritic shapes computed from the level set method
and phase-field model for $\Delta=0.55$, shown at time=9400.}
\end{figure}

The results presented here so far have used $D_S=D_L$, where $D_S$ and
$D_L$ are the diffusivities in the solid and liquid phases, respectively. 
With our level set algorithm, we can also investigate the more general
case where the diffusivities are unequal. We performed additional
simulations at $\Delta=0.65$ with $D_S=0.75,0.5,0.25$ and $0$ while
keeping $D_L=1$.  The only available benchmark for the case of
non-symmetric diffusion is the linearized solvability theory of Barbieri
and Langer \cite{Bar89}, which predicts
\begin{equation}
\rho^2V \approx \frac{1+D_S/D_L}{2}(\rho^2V)_{_{D_S/D_L=1}} 
\label{barbierilanger}
\end{equation}
where $\rho$ and $V$ are the steady-state tip radius and velocity, respectively. 
The values of $\rho^2V$ obtained from the level set simulations are compared 
with Eq.~6 in Figure 3. The fit is surprisingly good (an error of about $13\%$ 
at $D_S/D_L=0$), considering Eq.~6 was obtained from a linearized theory in the 
limit of small undercoolings.

\begin{figure}[tbh]
\leavevmode\centering\psfig{file=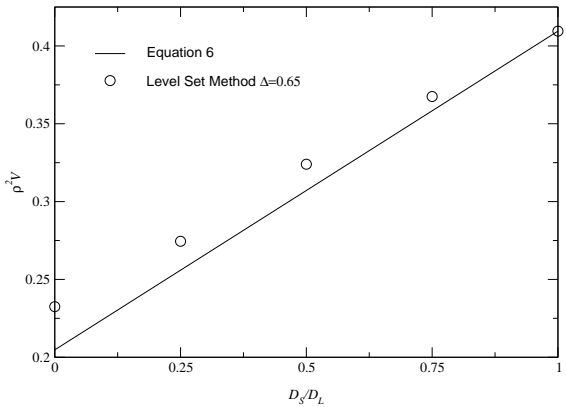,width=\columnwidth,angle=-90}
\caption{$\rho^2V$ for different values of $D_S/D_L$. The circles are data from 
level set simulations at $\Delta = 0.65$. The solid line is the theoretical 
prediction of Barbieri and Langer fitted to the data point at $D_S/D_L=1$.}
\end{figure}

In conclusion, the level set method should be considered as a viable
alternative to the use of phase-field models. We have used a level set
algorithm that can produce accurate calculations of dendritic growth which
can be compared favorably with solvability theory as well as
time-dependent phase-field model simulations. The level set method can
also handle discontinuous material properties easily, which is currently
very difficult with the phase-field approach. However, we should note that
our implementation is not at all efficient. The practical application of
this method to more realistic systems will require some sort of adaptive
technique. In the future, we would like to use more computationally
efficient implementations of this algorithm and apply these methods to
problems in directional solidification, where the ability to simulate
unequal diffusivities is of great interest.

We thank Susan Chen and Stanley Osher for useful discussions, Nikolas
Provatas for helpful remarks and for providing the adaptive phase-field
code used in this work, and Wouter-Jan Rappel for providing the
solvability code used to test our simulations. This work has been
supported by the NASA Microgravity Research Program, under Grant
NAG8-1249.  We also acknowledge the support of the National Center for
Supercomputing Applications (NCSA) for the use of its computer resources.

\bibliography{biblio}

\end{document}